\documentclass[doublecol]{epl2} 
\usepackage{graphicx}
\usepackage{epsfig}

\newcommand{\avg}[1]{\langle {#1} \rangle}

\title{Discrete-time Markov chain approach to contact-based disease spreading in complex networks}
\shorttitle{Contact-based epidemic spreading in complex networks}

\author{S. G\'omez\inst{1}, A. Arenas\inst{1}, J. Borge-Holthoefer\inst{1} S. Meloni\inst{2,3} \and Y. Moreno\inst{3,4}}

\institute{                    
  \inst{1} Departament d'Enginyeria Inform\`atica i Matem\`atiques, Universitat Rovira i Virgili, 43007 Tarragona, Catalonia, Spain\\
  \inst{2} Department of Informatics and Automation, University of Rome "Roma Tre", Via della Vasca Navale, 79 Rome 00146, Italy\\
  \inst{3} Instituto de Biocomputaci\'on y F\'\i sica de Sistemas Complejos (BIFI), Universidad de Zaragoza, Corona de Arag\'on 42, 50009 Zaragoza, Spain\\
  \inst{4} Department of Theoretical Physics, University of Zaragoza, 50009 Zaragoza, Spain
}


\abstract{
Many epidemic processes in networks spread by stochastic contacts among their connected vertices. There are two limiting cases widely analyzed in the physics literature, the so-called contact process (CP) where the contagion is expanded at a certain rate from an infected vertex to one neighbor at a time, and the reactive process (RP) in which an infected individual effectively contacts all its neighbors to expand the epidemics. However, a more realistic scenario is obtained from the interpolation between these two cases, considering a certain number of stochastic contacts per unit time. Here we propose a discrete-time formulation of the problem of contact-based epidemic spreading. We resolve a family of models, parameterized by the number of stochastic contact trials per unit time, that range from the CP to the RP. In contrast to the common heterogeneous mean-field approach, we focus on the probability of infection of individual nodes. Using this formulation, we can construct the whole phase diagram of the different infection models and determine their critical properties. 
}

\begin{document}

\maketitle

The problem of modeling how diseases spread among individuals has been intensively studied for many years\cite{hethcote,Maybook,daley,Murray}. The development of mathematical models to guide our understanding of the disease dynamics has allowed to address important issues such as immunization and vaccination policies\cite{Maybook,geisel,egamstw04}. Physicist's approaches to problems in epidemiology involve statistical physics, the theory of phase transitions and critical phenomena\cite{stanley}, which have been extremely helpful to grasp the macroscopic behavior of epidemic outbreaks\cite{vespiromu,vrpre,llm01,mpsv02,n02,bbpsv04,glmp08,nostre,cwwlf08}. The main artifice of this success has been the Mean-Field (MF) approximation, where local homogeneities of the ensemble are used to average the system, reducing degrees of freedom.

The study of complex networks\cite{newmanrev,yamirrep,dgm08} has provided new grounds to the understanding of contagion dynamics. Particularly important in nature are scale-free (SF) networks, whose degree distribution follows a power law $P(k)\sim k^{-\gamma}$ for the number of connections, $k$, an individual has. SF networks include patterns of sexual contacts\cite{stanley_nat}, the Internet\cite{vespbook08}, as well as other social, technological and biological networks\cite{caldarelli}. The critical properties of an epidemic outbreak in SF networks can be addressed using the heterogeneous MF (HMF) prescription\cite{vespiromu,vrpre,llm01,mpsv02,n02,bbpsv04,glmp08,nostre}. It consists of coarse-grained vertices within degree classes and considers that all nodes in a degree class have the same dynamical properties; the approach also assumes that fluctuations can be neglected. Specifically, if $\beta$ is the rate (probability per unit time) at which the disease spreads, it follows that the epidemic threshold in uncorrelated SF networks is given\cite{vespiromu} by $\beta_c=\langle k \rangle/\langle k^2 \rangle$, leading to $\beta_c \rightarrow 0$ as $N\rightarrow \infty$ when $2< \gamma \le 3$.

MF approaches are extremely useful to assess the critical properties of epidemic models, however they are not designed to give information about the probability of individual nodes but about classes of nodes. Then, questions concerning the probability that a given node be infected are not well posed in this framework. To obtain more details at the individual level of description, one has to rely on Monte Carlo (MC) simulations, which have also been used to validate the results obtained using MF methods. Restricting the scope of epidemiological models to those based in two states\cite{hethcote,daley,Murray} $-$susceptible (S) and infected (I)$-$, the current theory concentrates on two specific situations, the contact process\cite{Marro,Castellano:2006,HHP,Castellano:2007,Castellano:2008,marian} (CP) and the reactive process\cite{GA04,cbp05,cpv07} (RP). A CP stands for a dynamical process that involves an individual stochastic contagion per infected node per unit time, while in the RP there are as many stochastic contagions per unit time as neighbors a node has. This latter process underlies the abstraction of the susceptible-infected-susceptible (SIS) model\cite{hethcote,daley,Murray}. However, in real situations, the number of stochastic contacts per unit time is surely a variable of the problem itself \cite{nostre}. 

In this work, we introduce a theoretical framework for contact-based spreading of diseases in complex networks. Our formulation is based on probabilistic discrete-time Markov chains, generalizes existing HMF approaches and applies to weighted and unweighted complex networks. Within this context, in addition to capturing the global dynamics of the different contact models and its associated critical behavior, it is now possible to quantify the \emph{microscopic dynamics} at the individual level by computing the probability that any node is infected in the asymptotic regime. 
MC simulations corroborate that the formalism introduced here reproduces correctly the \emph {whole} phase diagram for model and real-world networks. Moreover, we capitalize on the new approach to address how the spreading dynamics depends on the number of contacts actually used by a node to propagate the disease. 

\section{Contact-based epidemic spreading models}

Let us consider a network made up of $N$ nodes, whose connections are represented by the entries $\{a_{ij}\}$ of an $N$-by-$N$ adjacency matrix ${\bf A}$. Additionally, in the most general case in which the network is weighted, we denote by $\{\omega_{ij}\}$ the weights of the connections between nodes, being $w_i=\sum_j w_{ij}$ the total strength\cite{barrat04a} of node $i$. The above quantities completely define the structure of the underlying graph. As for the dynamics, we consider a discrete two-state (S and I) contact-based process.
Each node of the network represents an individual (or a place, a city or airport for example) and each edge is a connection along which the infection spreads. At each time step, an infected node makes a number $\lambda$ of trials to transmit the disease to its neighbors with probability $\beta$ per unit time. This forms a Markov chain where the probability of a node being infected depends only on the last time step. After some transient time, the previous dynamics sets the system into a stationary state in which the average density of infected individuals, $\rho$, defines the prevalence of the disease.

We next look at the probability that any given node $i$ is infected at the stationary state. We denote by $r_{ij}$ the probability that a node $i$ is in contact with a node $j$, defining a matrix ${\bf R}$. These entries represent the probabilities that existing links in the network are used to transmit the infection. If $i$ and $j$ are not connected, then $r_{ij}=0$. Besides, $\mu$ stands for the rate at which infected nodes are recovered and get back to the susceptible class; and finally, $p_i(t)$ is the probability that a node $i$ is infected at time $t$. With these definitions, the discrete-time version of the evolution of the probability of infection of any node $i$ reads
\begin{eqnarray}
p_i(t+1) & = & (1-q_i(t))(1-p_i(t)) + (1-\mu) p_i(t) \nonumber\\
  & & \mbox{} + \mu (1-q_i(t)) p_i(t)
\label{pidet1}
\end{eqnarray}
where $q_i(t)$ is the probability of node $i$ not being infected by any neighbor
\begin{equation}
q_i(t) = \prod_{j=1}^{N} (1-\beta r_{ji} p_j(t))
\label{capin}
\end{equation}
The first term on the right hand side of eq.~(\ref{pidet1}) is the probability that node $i$ is susceptible ($1-p_i(t)$) and is infected ($1-q_i(t)$) by at least a neighbor. The second term stands for the probability that node $i$ is infected at time $t$ and does not recover, and finally the last term takes into account the probability that an infected node recovers ($\mu p_i(t)$) but is re-infected by at least a neighbor ($1-q_i(t)$). Within this formulation, we are assuming the most general situation in which recovery and infection occur on the same time scales, allowing then reinfection of individuals during a discrete time window (for instance, one MC step). This formulation generalizes previous approximations where reinfections can not occur.

The formulation so far relies on the assumption that the probabilities of being infected $p_i$ are independent random variables. This hypothesis turns out to be valid in the vast majority of complex networks because the inherent topological disorder makes dynamical correlations not persistent. The dynamical system (\ref{pidet1}, \ref{capin}) corresponds to a family of possible models, parameterized by the explicit form of the contact probabilities $r_{ij}$. Without loss of generality, it is instructive to think of these probabilities as the transition probabilities of random walkers on the network. The general case is represented by $\lambda_i$ random walkers leaving node $i$ at each time step:
\begin{equation}
r_{ij} = 1- \left( 1- \frac{w_{ij}}{w_i} \right)^{\lambda_i}
\label{mcon}
\end{equation}
The CP corresponds to a model dynamics of one contact per unit time, $\lambda_i=1$, $\forall i$ in eq.~(\ref{mcon}) thus $r_{ij} = w_{ij}/w_i$\footnote{Strictly speaking, when $\lambda=1$, our model is not exactly the standard CP, since there reinfections are not considered. However, we will refer to it as a CP since only one neighbor is contacted at each time step and the critical points of both variants are the same.}. In the RP all neighbors are contacted, which corresponds, in this description, to set the limit $\lambda_i \rightarrow \infty$,  $\forall i$ resulting on $r_{ij} = a_{ij}$ regardless of whether the network is weighted or not. Other prescriptions for $\lambda_i$ conform the spectrum of models that can be obtained using this unified framework.
The phase diagram of every model is simply obtained solving the system formed by eq.~(\ref{pidet1}) for $i=1,\ldots,N$ at the stationary state
\begin{equation}
p_{i} = (1-q_{i}) + (1-\mu) p_{i} q_{i}
\label{pista}
\end{equation}
This equation has always the trivial solution $p_i=0$, $\forall i=1,\ldots,N$. Other non-trivial solutions are reflected as non zero fixed points of eq.~(\ref{pista}) and can be easily computed numerically by iteration. The macroscopic order parameter is given by the expected infection density $\rho$, computed as
\begin{equation}
\rho = \frac{1}{N}   \sum_{i=1}^{N} p_{i}
\label{op}
\end{equation}

\section{Numerical results}

\begin{figure}[t]
  \begin{center}
  \begin{tabular}{l}
    \mbox{\includegraphics*[width=.45\textwidth]{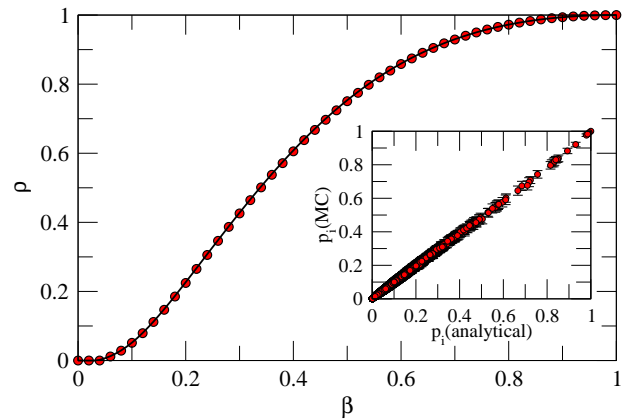}}
  \end{tabular}
  \end{center}
\caption{
Average fraction of infected individuals, $\rho$, as a function of the spreading rate $\beta$ for $N=10^4$. The symbols correspond to MC simulations of the SIS model on top of random scale-free networks with $\gamma=2.7$ (error bars are smaller than the size of the symbol) and the lines stand for the analytical solutions of our formalism (with $\lambda=\infty$).We also represent in the inset a scatter plot for the probability that a node $i$ ($i=1,\ldots,N$) is infected using results of MC simulations (the y-axis) and the solutions (x-axis) of eq.~(\ref{pista}). Both results have been obtained for $\mu=1$, the inset is for $\beta=0.1$.}
\label{fig1}
\end{figure}

To show the validity of the approach here discussed, we have performed MC simulations on different SF networks for RP. Figure~\ref{fig1} shows a comparison of the phase diagram of the system obtained by MC simulations, with the numerical solution of eq.~(\ref{pista}). To model the epidemic dynamics on the described topologies we incorporate a SIS model in which, at each time step, each node can be susceptible or infected. In our simulations time is discretized in time-steps and each simulation starts with a fraction $\rho_0$ of randomly chosen infected individuals ($\rho_0=0.05$ in our simulations). At each time step an infected node $i$ infects with the same probability $\beta$ all its neighbors and recovers at a rate $\mu$. The simulation runs until a stationary state for the density of susceptible individuals, $\rho(t)$ is reached. The agreement between both curves is matchless. 
Moreover, the formalism also captures the microscopic dynamics as given by the $p_i$'s, see the inset of fig.~\ref{fig1}.

In Figure~2 we analyze our formalism on top of the airports network data set, composed of passenger flights operating in the time period November 1, 2000, to October 31, 2001 compiled by OAG Worldwide (Downers Grove, IL) and analyzed previously by Prof.\ Amaral's group\cite{rogerair}. It consists of 3618 nodes (airports) and 14142 links, we used the weighted network in our analysis. Airports corresponding to a metropolitan area have been collapsed into one node in the original database. We show the density of infected individuals $\rho$ as a function of $\beta$ for different values of $\lambda$. The critical points as well as the shape of the $\rho-\beta$ phase diagrams greatly change at varying the number of stochastic contacts ($\lambda$). For small values of $\lambda$ the disease prevalence is moderate, even for large values of the spreading rate $\beta$. In contrast, when the number of trials is of order $10^3$ the situation is akin to a RP. 
\begin{figure}[t]
  \begin{center}
  \begin{tabular}{l}
    \mbox{\includegraphics*[width=.45\textwidth]{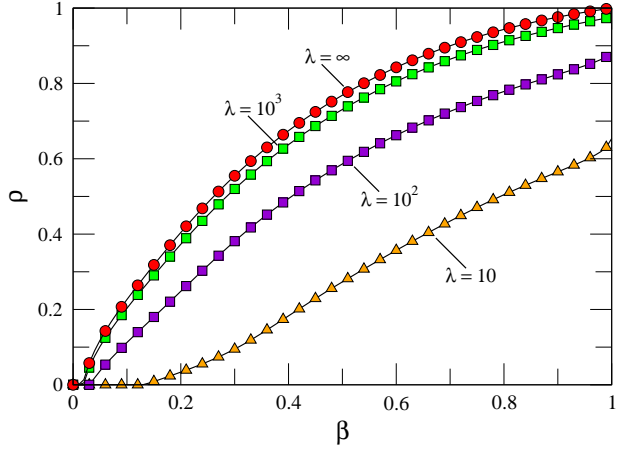}}
  \end{tabular}
  \end{center}
\caption{
Density of infected individuals $\rho$ as a function of $\beta$ for different values of $\lambda$ in the air transportation network \cite{rogerair}. 
We have set $\mu=1$ and $\rho$ is calculated according to eq.~(\ref{op}) once the $p_i$'s are obtained.}
\label{fig2}
\end{figure}

\begin{figure}
\begin{center}
  \begin{tabular}{l}
    \mbox{\includegraphics*[width=.45\textwidth]{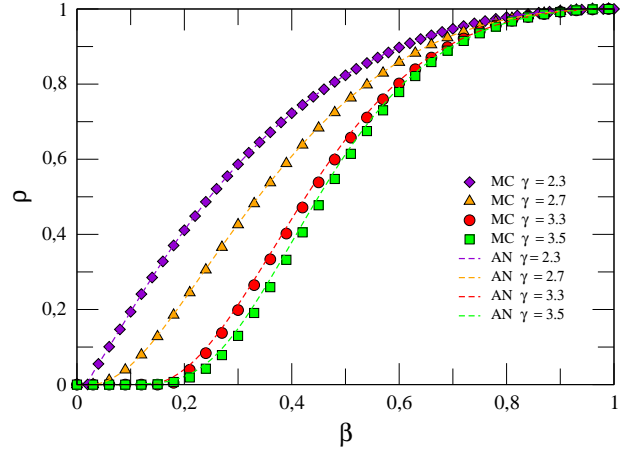}}
  \end{tabular}
  \end{center}
\caption{Phase diagram for the SIS model ($\lambda=\infty$) in a random scale free network for different $\gamma$'s. The networks are made up of $N=10^4$ nodes and $\mu=1$. MC results are averages over $10^2$ realizations. Dashed lines corresponds to the theoretical prediction and symbols to MC results.}
\label{fig3}
\end {figure}

\begin{figure}
\begin{center}
  \begin{tabular}{l}
    \mbox{\includegraphics*[width=.45\textwidth]{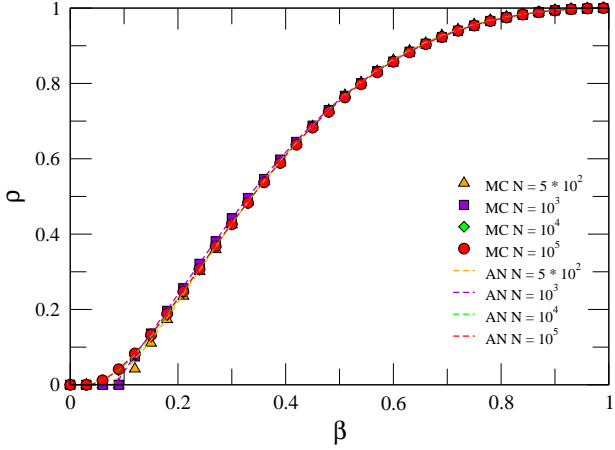}}
  \end{tabular}
  \end{center}
\caption{Phase diagram for the SIS model ($\lambda=\infty$) in a random scale free network for different system sizes as indicated. The networks have a power-law degree distribution with an exponent $\gamma=2.7$ and $\mu=1$. MC results are averages over $10^2$ realizations.}
\label{fig4}
\end {figure}

Finally, we compare the results of the formalism for different random scale-free networks satisfying $P(k)\sim k^{-\gamma}$ generated using the configuration model. Figure~\ref{fig3} shows the phase diagram for $\mu=1$ and several values of the exponent $\gamma$, both below and above $\gamma=3$. The system size has been fixed to $N=10^4$ nodes. The dotted lines represent the results obtained using the analytical approximation while symbols stand for MC simulations. As it can be seen, the agreement between both methods is remarkable, even for values of $\gamma<2.5$ where structural changes are extremely relevant \cite{shao}. The same agreement between MC results and the analytical solutions is obtained if one instead fixes the degree distribution exponent $\gamma$ and explores the dependency with the system size. This is what is shown in fig.~\ref{fig4}, where we have depicted the phase diagram for networks with $\gamma=2.7$ for several system sizes ranging from $N=500$ to $N=10^5$. Except for $N=500$, where MC results have a large standard deviation close to the critical point, the agreement is again excellent in the whole range of $\beta$ values.

\section{Epidemic Threshold}

Let us now assume the existence of a critical point $\beta_c$ for fixed values of $\mu$ and $\lambda_i$ such that $\rho=0$ if $\beta < \beta_c$ and $\rho>0$ when $\beta > \beta_c$. The calculation of this critical point is performed by  considering that when $\beta \rightarrow \beta_c$, the probabilities $p_i \approx \epsilon_i$, where $0\leq\epsilon_i \ll 1$, and then after substitution in eq.~(\ref{capin}) one gets
\begin{equation}
q_{i} \approx 1 - \beta \sum_{j=1}^{N} r_{ji} \epsilon_{j}
\label{qiap}
\end{equation}
Inserting eq.~(\ref{qiap}) in eq.~(\ref{pista}), and neglecting second order terms in $\epsilon$ we get

\begin{equation}
\sum_{j=1}^{N} \left( r_{ji} - \frac{\mu}{\beta} \delta_{ji} \right ) \epsilon_j = 0 \hspace{1cm} \forall i= 1,\ldots,N
\label{final}
\end{equation}

\noindent where $\delta_{ij}$ stands for the Kronecker delta. The system (\ref{final}) has non trivial solutions if and only if $\mu/\beta$ is an eigenvalue of the matrix ${\bf R}$. Since we are looking for the onset of the epidemic, the lowest value of $\beta$ satisfying (\ref{final}) is
\begin{equation}
\beta_c=\frac{\mu}{\Lambda_{\mbox{\scriptsize max}}}
\label{betac}
\end{equation}
where $\Lambda_{\mbox{\scriptsize max}}$ is the largest eigenvalue of the matrix ${\bf R}$. Equation~(\ref{betac}) defines the epidemic threshold of the disease spreading process. 

It is worth analyzing the two limiting cases of CP and RP above. In the first case, one obtains the trivial result that the only non-zero solution corresponds to $\beta_c=\mu$, because the matrix ${\bf R}$ is a transition matrix whose maximum eigenvalue is always ${\Lambda_{\mbox{\scriptsize max}}}=1$. For the RP corresponding to the SIS spreading process usually adopted\cite{vespiromu}, the classical result for uncorrelated SF networks is recovered because, in this case, the largest eigenvalue\cite{chungPNAS,restrepo} is $\Lambda_{\mbox{\scriptsize max}}=\langle k^2 \rangle / \langle k \rangle$.

\section{Mesoscopic equations at the critical point}

Once the general framework given by the dynamical system (\ref{pidet1}, \ref{capin}) has been proposed, it is instructive to approximate it using the hypotheses underlying HMF. These hypotheses consist of: i) coarse-graining the system in classes of node by degree, assuming that the dynamical properties within each class are the same, and ii) neglecting fluctuations. To obtain the mesoscopic description we consider the second order approximation of eq.~(\ref{pista}) and proceed as in the previous section. Therefore,
\begin{equation}
  q_i \approx 1 - \beta \sum_j r_{ji} \epsilon_j +  \beta^2 \sum_{j<l} r_{ji} r_{li} \epsilon_j \epsilon_l
\end{equation}
After substitution in (\ref{pista}) and reordering terms one gets
\begin{eqnarray}
  0 & = &  -\mu \epsilon_i + \beta (1 - \epsilon_i) \sum_j r_{ji} \epsilon_j +
    \mu \beta \epsilon_i \sum_j r_{ji} \epsilon_j \nonumber\\
    & & \mbox{} - \beta^2 \sum_{j<l} r_{ji} r_{li} \epsilon_j \epsilon_l
\label{tohmf}   
\end{eqnarray}
which are the equations governing the dynamics of the contact-based epidemic spreading process at the microscopic level. It is possible to write eq.~(\ref{tohmf}) at the commonly used mesoscopic (degree class) level for unweighted, undirected heterogeneous networks. The interactions then takes place between classes of nodes. Defining the average density of infected nodes with degree $k$ as $\rho_k=\frac{1}{N_k}\sum_{k_{i}=k} p_{i}$, where $N_k$ is the number of nodes with degree $k$ and the sum runs over the set of nodes of degree $k$, we obtain the generalized HMF equation near criticality.

\subsection{Homogeneous networks}
For homogeneous unweighted undirected networks,
$\epsilon_i = \epsilon$ and $k_i \approx \avg{k}$ for all nodes. Thus,
$\rho = \frac{1}{N} \sum_j \epsilon_j = \epsilon$ and
\begin{eqnarray}
  0 & = & -\mu \rho + \beta \rho (1 - \rho) \sum_j r_{ji} + \mu \beta \rho^2 \sum_j r_{ji} \nonumber\\
    & & \mbox{} - \beta^2 \rho^2 \sum_{j<l} r_{ji} r_{li}
  \label{meanfieldhomorji}
\end{eqnarray}
Defining
\begin{equation}
  R_{\lambda}(x)=1-(1 - x)^{\lambda}
\end{equation}
the terms involving values of $r_{ji}$ are
\begin{eqnarray}
  r_{ji} & \approx & a_{ji} R_{\lambda}(\avg{k}^{-1})
  \\
  \sum_j r_{ji} & \approx & \avg{k} R_{\lambda}(\avg{k}^{-1})
  \\
  \sum_{j<l} r_{ji} r_{li} & \approx & \frac{1}{2} \avg{k} (\avg{k} - 1) R_{\lambda}(\avg{k}^{-1})^2
\end{eqnarray}
Now, eq.~(\ref{meanfieldhomorji}) becomes
\begin{eqnarray}
  0 & = & -\mu \rho + \beta \rho (1 - \rho) \avg{k} R_{\lambda}(\avg{k}^{-1}) \nonumber \\
    &   & \mbox{} + \mu \beta \rho^2 \avg{k} R_{\lambda}(\avg{k}^{-1}) \nonumber \\
    &   & \mbox{} - \beta^2 \rho^2 \frac{1}{2} \avg{k} (\avg{k} - 1) R_{\lambda}(\avg{k}^{-1})^2
  \label{meanfieldhomo}
\end{eqnarray}
which may be considered as the MF approximation of our model for homogeneous networks.

If $\lambda=1$ then $R_{1}(\avg{k}^{-1}) = \frac{1}{\avg{k}}$
and eq.~(\ref{meanfieldhomo}) becomes
\begin{equation}
0 = -\mu \rho + \beta \rho (1 - \rho) + \mu \beta \rho^2 - \frac{\avg{k}-1}{2\avg{k}} \beta^2 \rho^2
\end{equation}

If $\lambda \rightarrow \infty$ then $R_{\infty}(\avg{k}^{-1}) = 1$
and eq.~(\ref{meanfieldhomo}) reads
\begin{equation}
  0 = -\mu \rho + \beta \rho (1 - \rho) \avg{k} +
    \mu \beta \rho^2 \avg{k} -
    \frac{1}{2} \beta^2 \rho^2 \avg{k} (\avg{k} - 1)
\end{equation}

In both cases, the first two terms correspond to the standard CP and RP models
(previously reported in the literature) respectively, and the additional terms
are second order contributions corresponding to reinfections and multiple infections.

\subsection{Heterogeneous networks}

Now we will concentrate on the class of heterogeneous unweighted undirected networks
completely specified by their degree distribution $P(k)$ and by the conditional
probability $P(k'|k)$ that a node of degree $k$ is connected to a node of degree $k'$.
Of course, the normalization conditions $\sum_k P(k) = 1$ and $\sum_{k'} P(k'|k) = 1$
must be fulfilled. In this case, the average number of links that goes from a node
of degree $k$ to nodes of degree $k'$ is $k P(k'|k)$.

In these heterogeneous networks it is supposed that all nodes of the same degree behave
equally, thus $\epsilon_i = \epsilon_j$ if $k_i=k_j$, and the density $\rho_k$ of infected
nodes of degree $k$ is given by
$\rho_k = \frac{1}{N_k} \sum_{i\in K} \epsilon_i = \epsilon_j\,,\ \ \forall j\in K$,
where $N_k = P(k) N$ is the expected number of nodes with degree $k$. Here we have
made use of $K$ to denote the set of nodes with degree $k$. This notation allows
to group the sums by the degrees of the nodes,
for instance
\begin{equation}
  \sum_j a_{ji} R_{\lambda}(k_j^{-1}) \epsilon_j =
    k \sum_{k'} P(k'|k) R_{\lambda}({k'}^{-1}) \rho_{k'}
\end{equation}
After some algebra eq.~(\ref{tohmf}) leads to the generalized HMF equation
\begin{eqnarray}
 && 0  =  -\mu \rho_{k}
          + \beta k (1 - \rho_{k}) \sum_{k'} P(k'|k) R_{\lambda}({k'}^{-1}) \rho_{k'}
          \nonumber \\
    &   & \mbox{} + \mu \beta k \rho_{k} \sum_{k'} P(k'|k) R_{\lambda}({k'}^{-1}) \rho_{k'}
          \nonumber \\
    &   & \mbox{} + \frac{1}{2}\beta^2 k \sum_{k'} R_{\lambda}({k'}^{-1})^2 P(k'|k) \rho_{k'}^2
          \nonumber \\
    &   & \mbox{} - \frac{1}{2}\beta^2 k^2 \left( \sum_{k'} R_{\lambda}({k'}^{-1}) P(k'|k) \rho_{k'} \right)^2
  \label{mfhetero2}
\end{eqnarray}
If $\lambda=1$ then $R_{1}(k^{-1}) = \frac{1}{k}$
and eq.~(\ref{mfhetero2}) becomes
\begin{eqnarray}
 && 0 =   -\mu \rho_{k}
          + \beta k (1 - \rho_{k}) \sum_{k'} \frac{1}{k'} P(k'|k) \rho_{k'}
          \nonumber \\
    &   & \mbox{} + \mu \beta k \rho_{k} \sum_{k'} \frac{1}{k'} P(k'|k) \rho_{k'}
                  + \frac{1}{2}\beta^2 k \sum_{k'} \frac{1}{{k'}^2} P(k'|k) \rho_{k'}^2
          \nonumber \\
    &   & \mbox{} - \frac{1}{2}\beta^2 k^2 \left( \sum_{k'} \frac{1}{k'} P(k'|k) \rho_{k'} \right)^2
\end{eqnarray}
If $\lambda \rightarrow \infty$ then $R_{\infty}(k^{-1}) = 1$
and eq.~(\ref{mfhetero2}) reads
\begin{eqnarray}
 && 0 =   -\mu \rho_{k}
          + \beta k (1 - \rho_{k}) \sum_{k'} P(k'|k) \rho_{k'}
          \nonumber \\
    &   & \mbox{} + \mu \beta k \rho_{k} \sum_{k'} P(k'|k) \rho_{k'}
                  + \frac{1}{2}\beta^2 k \sum_{k'} P(k'|k) \rho_{k'}^2
          \nonumber \\
    &   & \mbox{} - \frac{1}{2}\beta^2 k^2 \left( \sum_{k'} P(k'|k) \rho_{k'} \right)^2
\end{eqnarray}
Again, the first two terms in both cases correspond to the standard CP and RP
HMF equations respectively, and the additional terms
are second order contributions corresponding to reinfections and multiple infections.

\section{Conclusions}

We have proposed a new framework to study disease spreading in networks. By defining a set of discrete-time equations for the probability of individual nodes to be infected, we construct a dynamical system that generalizes from an individual contact process to the classical case in which all connections are concurrently used, for any complex topology. Solving the equations at the stationary state, we find the whole phase diagram of the system. The numerical solution of the analytic equations overcomes the computational cost of MC simulations. Moreover, the formalism allows to gain insight on the behavior of the critical epidemic threshold for different values of the probability of contacting a fraction $\lambda$ of neighbors per time step. The proposed model deals with infections driven by direct contacts between nodes, but not with traffic situations where nodes transmit the epidemics by flow communication with others \cite{nostre}. In this latter case, the routing protocol of traffic between nodes is absolutely relevant and can change the critical point of the epidemic spreading. We are currently working to adapt the present formalism also to traffic situations.


\end{document}